\documentclass[3p,times]{elsarticle}

\usepackage{ecrc}

\newcommand {\bea}{\begin{eqnarray}}
\newcommand {\eea}{\end{eqnarray}}
\newcommand {\nn}{\nonumber \\}
\newcommand {\pl}{\partial}

\newcommand {\al}{\alpha}
\newcommand {\be}{\beta}

\newcommand {\x}{\xi}

\newcommand {\om}{\omega}
\newcommand {\Om}{\Omega}

\newcommand {\e} {\mbox{e}}

\newcommand {\del}  {\delta}
\newcommand {\Del}  {\Delta}
\newcommand {\Dcal}{{\cal D}}

\newcommand {\vtil}{{\tilde v}}
\newcommand {\Xtil}{{\tilde X}}

\newcommand {\Vbar}  {{\bar V}}

\newcommand {\lbar}{\bar{\ell}}
\newcommand {\vbar}{\bar{v}}
\newcommand {\xbar}{\bar{x}}

\newcommand {\rdot}{\dot{r}}

\newcommand {\wdot}{\dot{w}}
\newcommand {\xdot}{\dot{x}}
\newcommand {\ydot}{\dot{y}}

\newcommand {\xddot}{\ddot{x}}

\newcommand {\etap}{{\eta'}}

\newcommand {\ra} {\rightarrow}

\newcommand {\pr}   {{\quad .}}
\newcommand {\com}  {{\quad ,}}
\newcommand {\q}    {\quad}

\newcommand {\nl}    {\newline}

\newcommand {\PL}   {Phys.Lett.}

\newcommand {\xn} {{x_n}}
\newcommand {\Xn} {{X_n}}
\newcommand {\Pn} {{P_n}}

\newcommand {\alinv} {{\frac{1}{\alpha}}}

\volume{00}

\firstpage{1}

\journalname{Tribology International}

\runauth{S. Ichinose}

\jid{article}

\jnltitlelogo{Tribology International}

\CopyrightLine{2015}{Published by Elsevier Ltd.}

\usepackage{amssymb}
\usepackage[figuresright]{rotating}

\begin{document}

\begin{frontmatter}



\dochead{Full length article}
%% e.g. \dochead{17th International Conference on Dynamical Processes in Excited States of Solids}

\title{Non-equilibrium Statistical Approach to Friction Models}


\author{Shoichi Ichinose}

\address{
Laboratory of Physics, School of Food and Nutritional Sciences, 
University of Shizuoka, 
Yada 52-1, Shizuoka 422-8526, Japan\\
Tel: 81-54-264-5224,  Fax: 81-54-264-5099, Email: ichinose@u-shizuoka-ken.ac.jp
            }

\begin{abstract}
A geometric approach to the friction phenomena is presented. 
It is based on the holographic view which has recently been popular 
in the theoretical physics community. We see the system in one-dimension-higher space. 
The heat-producing phenomena are most widely treated 
by using the non-equilibrium statistical physics. 
We take 2 models of the earthquake. 
The dissipative systems are here formulated 
from the geometric standpoint. The statistical fluctuation is taken into account by using 
the (generalized) Feynman's path-integral. 
\end{abstract}


\end{frontmatter}

% \linenumbers

\twocolumn
%\begin{multicols}{2}
%%%%%%%%%%%%%%%%%%%%%%%%%%%%%%%%%%%%%%%%%%%%%%%%%%%%%%%%%%%%%%%%%%%%%
%%%%%%%%%%%%%%%%%%%%%%%%%%%  Sec.1 Introduction %%%%%%%%%%%%%%%%%%%%%%%%%
%%%%%%%%%%%%%%%%%%%%%%%%%%%%%%%%%%%%%%%%%%%%%%%%%%%%%%%%%%%%%%%%%%%%%
%{\bf\large 1.\  Introduction}\q
\section{INTRODUCTION\label{intro}}
%***label**{intro}
%%%%%%%%%%%%%%
The system 
we consider consists of the huge number of particles (blocks) and 
the size of the constituent particles is the {\it mesoscopic}-scale. 
It is lager than 50 nm =$5\times 10^{-8}$m and is 
far bigger than the atomic scale ($\sim 10^{-10}$m).  It is smaller than or nearly 
equal to the optical microscope scale ($\sim 10^{-6}$m) in the branches such as the soft-matter physics, 
the nano-science physics and the biophysics. The larger end of the mesoscopic (length) scale 
depends on  each phenomenon. For the earthquake it is about $10^{-4}$m. 

The physical quantities, such as velocity, 
energy and entropy, are the {\it statistically-averaged} ones. 
It is not obtained by the deterministic way like the classical (Newton) mechanics. 
{\it Renormalization} phenomenon  occurs not from the quantum effect but from 
the statistical fluctuation due to 
the uncertainty caused by the following facts. 
Firstly each particle obeys the Newton's law with different 
{\it initial conditions}. 
The total number of particles, N, is so large that we do not or can not observe the initial data. Usually we do not have interest 
in the trajectory of every particle and do not observe it. We have interest only in the macroscopic 
quantities: {\it total energy} and {\it total entropy} are the most important ones. 
Secondly, in the real system, the size and shape differ particle by particle.  
We regard the randomness as a part of fluctuation. 
Finally the models, presented in the following, contains discrete parameters ($t_n$ in 
Sec.2 and $y_n$ in Sec.3).  As far as the discreteness is kept 
(in the case that we no not take the continuous limit), 
the quantities determined by the minimal principle include the inevitable ambiguity which is 
regarded as a part of fluctuation. 

After the development of the string and D-brane theories\cite{StringText1, StringText2}, 
one general relation, between the 4-dimensional(4D) conformal theories and the 5D gravitational 
theories, was proposed. The 5D gravitational theories are asymptotically AdS$_5$\cite{Malda9711,GKP9802,Witten9802}. 
The proposal claims the quantum behavior of the 4D theories is obtainable 
by the classical analysis of the 5D gravitational ones. The development along the extra axis can be 
regarded as the renormalization flow. This approach (called AdS/CFT) has been providing 
non-perturbative studies in several branches: quark-gluon plasma physics, heavy-ion collisions, 
non-equilibrium statistical mechanics, superconductivity, superfluidity\cite{Natsu2015, AE2015}. 
Especially, as the most relevant 
to the present work, the connection with the hydrodynamics is important\cite{Natsu08}. 
When a black hole is given a perturbation, the effect decays as the relaxation phenomenon. The transport coefficients, such as viscosities, speed of sound, 
thermal conductivity, are important physical quantities.

We take, in Sec.3,  Burridge-Knopoff model for the earthquake analysis\cite{RTSKS2003,Zion2008}. 
It was first introduced by Burridge and Knopoff\cite{BK1967}. 
Carlson, Langer  and collaborators performed a pionering study of the statistical properties 
\cite{CL1989a, CL1989b}. Further development was reviewed in ref.\cite{KHK2012}. 

We exploit 
the {\it computational step number} $n$ instead of (usual) time.  
The step flow is given by 
the discrete Morse flows theory\cite{Kikuchi, Kikuchi2}. 
In the first model (Sec.2), 
we adopt this step-wise approach for the {\it time}-development. 
The time variable is introduced as $t_n=nh~(h:\mbox{time-interval unit})$. 
In the second model (Sec.3), 
we take the approach for the {\it space}-propagation. 
The position variable is introduced as $y_n=na~(a:\mbox{space-interval unit})$. 
The non-equilibrium dissipative system is recently formulated using 
the discrete Morse flows theory combined with the (generalized) path-integral
\cite{SI1303Cam, SI1308APPC}.

%%%%%%%%%%%%%%%%%%%%%%%%%%  Sec.2  %%%%%%%%%%%%%%%%%%%%%%%%%%%%%%%%%
%%%%%%%%%%%%                2. Spring-Block Model                                %%%%%%
%%%%%%%%%%%%%%%%%%%%%%%%%%%%%%%%%%%%%%%%%%%%%%%%%%%%%%%%%%%%%%%%%%
%{\bf\large 2.\q SPRING-BLOCK MODEL}\q
\section{SPRING-BLOCK MODEL \label{SB}}
\begin{figure}[h]
                  %%  <minipage A
\begin{minipage}{16pc}
\includegraphics[width=16pc,bb= 0 0 297 130]{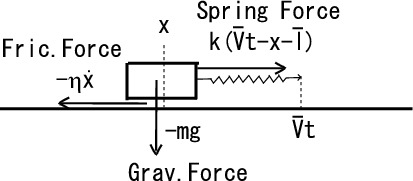}
\caption{
%****SBmodel.eps*** 
{\it The spring-block model, (\ref{SB2b}). }
        }
\label{SBmodel}
\end{minipage} 
                   %%  minipage A>
\hspace{3pc}
                   %%  <minipage B
\begin{minipage}{16pc}
\includegraphics[bb= 0 0 122 92]{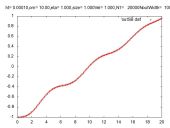}
\caption{
%****MovSB.eps***
{\it Spring-Block Model, Movement, } 
$h$=0.0001,$\sqrt{k/m}$=10.0, $\eta/m$=1.0, $\Vbar$=1.0, $\lbar$=1.0, total step no =20000. 
{\it The step-wise solution (\ref{SB2}) correctly reproduces the analytic solution: }
$x(t)=\e^{-\eta' t/2}\Vbar\{ ({\eta'}^2/2\om^2-1)(\sin\Om t)/\Om +(\eta'/\om^2)\cos\Om t \}
-\lbar+\Vbar (t-\eta'/\om^2)\com \Om=(1/2)\sqrt{4\om^2-{\eta'}^2}=9.99\com
\ 0\leq t\leq 2\com\q x(0)=-\lbar,\ \xdot(0)=0$. 
        }
\label{MovSB}
\end{minipage}
                   %%  minipage B
                %%%%%%%%%%%%%%%%%%%%%%      Fig.1 and 2>   %%%%%%%%%%%%%%%%
\end{figure}

%**** SB ****
  We treat the movement of a block which is pulled by the spring which moves 
at the constant speed $\Vbar$. The block 
moves on the surface with friction. 
This is called the spring-block (SB) model. 
We adopt the {\it discrete Morse flows} method to treat this non-equilibrium system
\cite{Kikuchi, Kikuchi2}. 
We take the following $n$-th energy function 
to define the step($n$) flow. 
%*** SB1%%%%%%%%%%%%%%%%
\bea
K_n(x)=  V(x)-hnk\Vbar x +
\frac{\eta}{2h}(x-x_{n-1})^2\nn
+\frac{m}{2h^2}(x-2x_{n-1}
+x_{n-2})^2+K_n^0\ ,\ 
V(x)=\frac{kx^2}{2}+k\lbar x
,
\label{SB1}
\eea 
%%%%%%%%%%%%%%%%%%%%%%%%%%%
where  $\eta$ is the friction coefficient and $m$ is the block mass. 
$h$ is the 'time' interval parameter. 
$x$ is the position of the block. 
The potential $V(x)$ has two terms: one is the harmonic oscillator
with the spring constant $k$, and the other is the linear term of x with 
a new parameter $\lbar$ (the natural length of the spring). 
$\Vbar$ is the velocity (constant) with which the front-end of the spring moves. 
$K_n^0$ is a constant which does not depend on $x$. % It will be fixed later. 
The $n$-th step $\xn$ is determined by the energy minimum principle: 
$\del K_n(x)|_{x=\xn}=0$ with the pre-known position at the (n-1)-th, $x_{n-1}$, 
and that at the (n-2)-th, $x_{n-2}$. 
%*** SB2%%%%%%%%%%%%%%%%
\bea
\frac{k}{m}(\xn+\lbar-nh\Vbar)+
\frac{1}{h^2}(\xn-2x_{n-1}+x_{n-2})+\nn
\frac{\eta}{m}\frac{1}{h}(\xn-x_{n-1})=0\ ,\ 
\om\equiv \sqrt{\frac{k}{m}}\ ,\ \etap\equiv \frac{\eta}{m}, 
\label{SB2}
\eea 
%%%%%%%%%%%%%%%%%%%%%%%%%%%
where $n=2,3,4,\cdots$. 
For the continuous {\it time} limit: $h\ra 0, nh=t_n\ra t, v_n\equiv (\xn-x_{n-1})/h\ra\xdot, 
(\xn-2x_{n-1}+x_{n-2})/h^2\ra\xddot$, the above recursion relation reduces to 
the following differential equation. 
%*** SB2b%%%%%%%%%%%%%%%%
\bea
m\xddot=k(\Vbar t-x-\lbar)-\eta\xdot
\pr
\label{SB2b}
\eea 
%%%%%%%%%%%%%%%%%%%%%%%%%%%
This is the ordinary one for the spring-block model. See Fig.\ref{SBmodel}.

\begin{figure}[h]
                  %%  <minipage A
\begin{minipage}{16pc}
\includegraphics[bb= 0 0 122 92]{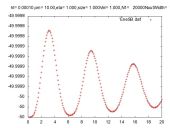}
\caption{
%****EneSB.eps***
{\it Spring-Block Model, Energy Change, }
$h$=0.0001,$\sqrt{k/m}$=10.0, $\eta/m$=1.0, $\Vbar$=1.0, $\lbar$=1.0, {\it total step no} =20000.
        }
\label{EneSB}
\end{minipage} 
                   %%  minipage A>
\hspace{3pc}
                   %%  <minipage B
\begin{minipage}{16pc}
\includegraphics[width=10pc, bb= 0 0 376 309]{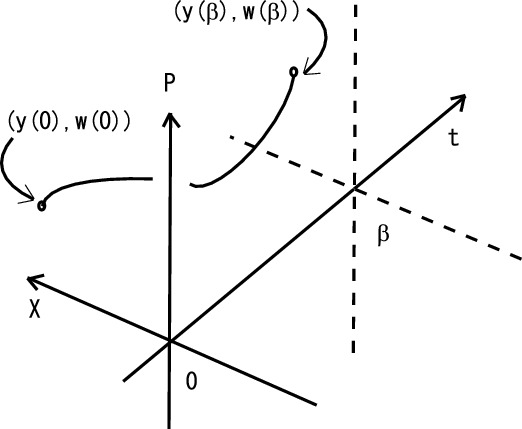}
\caption{
%****LinePath.eps*** 
{\it The path} $\{(y(t),w(t),t) | 0\leq t\leq \be  \}$ {\it of line in 3D bulk space} (X,P,t). 
        }
\label{LinePath}
\end{minipage}
                   %%  minipage B
\end{figure}

The graph of movement ($\xn$, eq.(\ref{SB2})) is shown in Fig.\ref{MovSB}. From the graph, 
we see this system 
starts with the {\it stick-slip} motion and 
reaches the {\it steady state} as $n\ra \infty$. 
Fig.\ref{EneSB} shows the energy change as the step flows. 
It shows the energy oscillates periodically and the amplitude 
goes down as the step goes. 
The physical dimensions of the 
parameters in (\ref{SB2b}) are listed as
%*** SBp%%%%%%%%%%%%%%%%
\bea
[m]=\mbox{M},[k]=\mbox{M}\mbox{T}^{-2},[\lbar]=\mbox{L}, 
\ [\eta]=\mbox{M}\mbox{T}^{-1},\nn
 { [{\Vbar}]}
=\mbox{L}\mbox{T}^{-1},
\label{SBp}
\eea 
%%%%%%%%%%%%%%%%%%%%%%%%%%%
where we assume $[x]=$L, $[t]=$T and $[h]=$T. 
(M:~mass, T:~time, L:~length.)

Now we consider N {\it copies} of the one body system (\ref{SB2}). N is 
sufficiently large, for example,  $10^{23}$(1 mol). 
We are {\it modeling} the present statistical system as follows. 
The N particles are "moderately" interacting each other in such way that 
each particle almost independently  moves except that energy is exchanged. 
The interaction is not so strong as to break the dynamics (\ref{SB2}). 
We use Feynman's path-integral method in order to 
take the statistical average of this N-copies system. The statistical 
ensemble measure will be given explicitly. 

From the energy expression (\ref{SB1}), we can read the {\it metric} ({\it geometry}) of 
this mechanical system.  
%*** SB11%%%%%%%%%%%%%%%%
\bea
{\Del s_n}^2\equiv 2h^2(K_n(\xn)-K_n^0)\nn 
=2~dt^2V_1(\Xn)+(\Del\Xn)^2 
+(\Del \Pn)^2,\nn
V_1(\Xn)\equiv V(\frac{\Xn}{\sqrt{\eta h}})-nk\sqrt{\frac{h}{\eta}}\Vbar\Xn,
\ dt\equiv h
,
\label{SB11}
\eea 
%%%%%%%%%%%%%%%%%%%%%%%%%%%
where $\Xn\equiv \sqrt{\eta h}\xn,~\Pn/\sqrt{m}\equiv h v_n=(\xn-x_{n-1}), $.
Using this metric, we can introduce the associated {\it statistical ensemble}  
of the spring-block model\cite{SI1010ICSF, SI1203ITChiro, SI1305WTCtorino}.  \nl

[Statistical Ensemble 1a]

The first choice of the metric in the 3D (t,X,P) manifold is the Dirac-type one:
%*** se1%%%%%%%%%%%%%%%%
\bea
(ds^2)_D\equiv 2V_1(X)dt^2+dX^2+dP^2\nn 
-\ \mbox{on-path}~(X=y(t), P=w(t))\ra \nn 
(2V_1(y)+\ydot^2+\wdot^2) dt^2,
\label{se1}
\eea 
%%%%%%%%%%%%%%%%%%%%%%%%%%%
where $\{(y(t),w(t))| 0\leq t\leq \be\}$ is a path (line) in the 3D space. 
See Fig.\ref{LinePath}. 
(The momentum variable $P$ is the "extra" coordinate in the holographic view. ) 
The length between $0\leq t_n=nh\leq \be$ and the 
ensemble measure are given by 
%*** se1b%%%%%%%%%%%%%%%%
\bea
L_D=\int_0^\be ds|_{on-path}=\int_0^\be\sqrt{2V_1(y)+\ydot^2+\wdot^2}dt \nn
=h \sum_{n=0}^{\be/h}\sqrt{2V_1(y_n)+\ydot_n^2+\wdot_n^2},\ 
d\mu=\e^{-\alinv L_D}\prod_t\Dcal y\Dcal w, \nn
\e^{-\be F}=\int\prod_n dy_n dw_n\e^{-\alinv L_D}
,
\label{se1b}
\eea 
%%%%%%%%%%%%%%%%%%%%%%%%%%%
where the free energy $F$ is defined. 
($\e^{-\be F}$ is the partition function.) 
$\al$ is a parameter ('string tension') with 
the dimension of $\sqrt{M}L/T$. 
The dimensionless one $\al'$ can be defined by $\al'=\al/D_c$ 
where $D_c$ is the characteristic dimensional unit with the dimension of $\sqrt{M}L/T$. 
For example $\sqrt{m}\lbar/h$, $\lbar\eta/\sqrt{m}$, 
or $\lbar\sqrt{k}$. $\al'$ is determined by the experimental data. \nl 

[Statistical Ensemble 1b]

The second choice of the metric is the standard type:  
%*** se2%%%%%%%%%%%%%%%%
\bea
(ds^2)_S\equiv \frac{1}{dt^2}[(ds^2)_D]^2\ \ -\mbox{on-path}\ra \nn 
(2V_1(y)+\ydot^2+\wdot^2)^2 dt^2.
\label{se2}
\eea 
%%%%%%%%%%%%%%%%%%%%%%%%%%%
Then the statistical ensemble is given by using the following length:
%*** se2b%%%%%%%%%%%%%%%%
\bea
L_S=\int_0^\be ds|_{on-path}=\int_0^\be (2V_1(y)+\ydot^2+\wdot^2)dt=\nn
h \sum_{n=0}^{\be/h}(2V_1(y_n)+\ydot_n^2+\wdot_n^2),\nn
d\mu=\e^{-\alinv L_S}\Dcal y\Dcal w,\ 
\e^{-\be F}=\int\prod_n dy_n dw_n\e^{-\alinv L_S}\nn 
=(\mbox{const})\int\prod_{n=0}^{\be/h}dy_n\e^{-\frac{h}{\al}(2V_1(y_n)+\ydot_n^2)}
.
\label{se2b}
\eea 
%%%%%%%%%%%%%%%%%%%%%%%%%%%
Note that $w(t)$ decouples from $y(t)$. 
The last expression is, when $\Vbar=0, \lbar=0$, the same as the partition function of the system of 
N($=\be/h$) harmonic oscillators with the frequency $\sqrt{k/\eta h}$\cite{SI1010ICSF}. 
The minimal path of (\ref{se2b}), by changing $y_n~\ra y,~nh~\ra t$ and using the variation 
$y~\ra~y+\del y$, we obtain
%*** SB2bb%%%%%%%%%%%%%%%%
\bea
-\eta h\xddot=k(\Vbar t-x-\lbar),\ \ x=\frac{y}{\sqrt{\eta h}}
\pr
\label{SB2bb}
\eea 
%%%%%%%%%%%%%%%%%%%%%%%%%%%
Comparing with (\ref{SB2b}), we notice 
1)~the viscous term disappeared;~
2)~the mass parameter $m$ is replaced by $\eta h$;~
3)~the sign in front of the acceleration-term (inertial-term) is different. 
By changing to the Euclidean time 
$\tau=i t$, the above equation reduces to the harmonic oscillator when we take 
$\Vbar=0,~\lbar=0$.   \nl

[Statistical Ensemble 2]

\begin{figure}[h]
                  %%  <minipage A
\begin{minipage}{16pc}
\includegraphics[width=8pc, bb= 0 0 357 310]{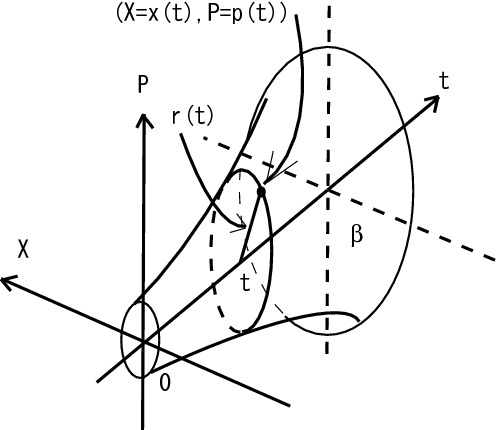}
\caption{
%****2DHyperSurf.eps*** 
{\it The two dimensional surface, (\ref{se3}),  in 3D bulk space (X,P,t). }
        }
\label{2DHyperSurf}
\end{minipage} 
                   %%  minipage A>
\hspace{3pc}
                   %%  <minipage B
\begin{minipage}{16pc}
\includegraphics[bb= 0 0 230 115]{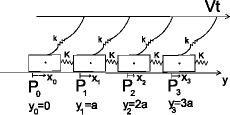}
\caption{
%****Fig6BK.eps*** 
{\it Burridge-Knopoff Model\  (\ref{BK2})}
        }
\label{Fig6BK}
\end{minipage}
                   %%  minipage B
\end{figure}

The first two statistical ensembles are based on the {\it line} in the 3D manifold (t,X,P). 
We present here another type which is based on the {\it surface} in the 3D space. 
First the 3D metric (Dirac type) is re-expressed in the following general form.  
%*** se1z%%%%%%%%%%%%%%%%
\bea
(ds^2)_D\equiv 2V_1(X)dt^2+dX^2+dP^2\equiv e_1 G_{IJ}(\Xtil)d\Xtil^I d\Xtil^J,\nn
I,J=0,1,2;\ (\Xtil^0,\Xtil^1,\Xtil^2)\equiv (t/d_0,X/d_1,P/d_2)\nn
e_1=m\lbar^2,\ d_0=\sqrt{\frac{k}{m}}, d_1=d_2=\sqrt{m}\lbar,\nn 
(G_{IJ})=\left(
\begin{array}{ccc}
2{d_0}^2V_1(d_1\Xtil^1)  &  0  &  0  \\
0 & {d_1}^2 &  0 \\
0 & 0 & {d_2}^2,
\end{array}
          \right)
\label{se1z}
\eea 
%%%%%%%%%%%%%%%%%%%%%%%%%%%
where we have introduced the {\it dimensionless} coordinates $\Xtil^I$. 
Here we introduce the following surface to define a "path" (surface) in the 3D space. 
%*** se3%%%%%%%%%%%%%%%%
\bea
\frac{X^2}{{d_1}^2}+\frac{P^2}{{d_2}^2}=\frac{r(t)^2}{{d_1}^2},\ \ \ 0\leq t\leq \be
,
\label{se3}
\eea 
%%%%%%%%%%%%%%%%%%%%%%%%%%%
where the radius parameter r is chosen to have the dimension of $\sqrt{M}L$. 
See Fig.\ref{2DHyperSurf}. Then we can define the {\it induced metric} $g_{ij}$ on the 2D surface.
%*** se4%%%%%%%%%%%%%%%%
\bea
\left.(ds^2)_D\right|_{\mbox{on-path}}=\left. 2V_1(X)dt^2+dX^2+dP^2\right|_{\mbox{on-path}}\nn
=e_1 \sum_{i,j=1}^{2}g_{ij}(\Xtil)d\Xtil^id\Xtil^j\com\q e_1=m\lbar^2\com\nn
(g_{ij})=\left(
\begin{array}{cc}
1+\frac{e_1}{{d_1}^2}\frac{2V_1}{r^2\rdot^2}X^2& \frac{e_1}{d_1d_2}\frac{2V_1}{r^2\rdot^2}X P \\
\frac{e_1}{d_1d_2}\frac{2V_1}{r^2\rdot^2}P X & 1+\frac{e_1}{{d_2}^2}\frac{2V_1}{r^2\rdot^2}P^2
\end{array}
          \right)
,
\label{se4}
\eea 
%%%%%%%%%%%%%%%%%%%%%%%%%%%
Using the (dimensionless) surface area $A$, the third partition function $\e^{-\be F}$ is given by
%*** se5%%%%%%%%%%%%%%%%
\bea
A=\int\sqrt{\det g_{ij}}d^2\Xtil=\frac{1}{d_1d_2}\int\sqrt{1+\frac{2V_1}{\rdot^2}}dX dP,\nn
\e^{-\be F}=\int_0^\infty d\rho\int_{\begin{array}{c}
                                                        r(0)=\rho \\
                                                        r(\be)=\rho
                                                 \end{array}}
\prod_t\Dcal X(t)\Dcal P(t)\e^{-\frac{1}{\al}A}
,
\label{se5}
\eea 
%%%%%%%%%%%%%%%%%%%%%%%%%%%
where $\al$ is the (dimensionless) "string" constant and here is a model 
parameter. It is determined by fitting the present result with the experimental data. 
(Note $\frac{2V_1}{\rdot^2}$ is dimensionless. )

%%%%%%%%%%%%%%%%%%%%%%%%%%%%  Sec.3  %%%%%%%%%%%%%%%%%%%%%%%%%%%%%%%%%
%%%%                Burridge-Knopoff Model                                    %%%%%%%%%%%%
%%%%%%%%%%%%%%%%%%%%%%%%%%%%%%%%%%%%%%%%%%%%%%%%%%%%%%%%%%%%%%%%%%%%%
%{\bf\large 3.\q Burridge-Knopoff Model}\q
\section{BURRIDGE-KNOPOFF MODEL  \label{BK}}
%***label***{BK}

 Let us take the following $n$-th energy function 
to define Burridge-Knopoff (BK) model in the step($n$) flow method. 
%*** BK1%%%%%%%%%%%%%%%%
\bea
I_n(x)=-xF(\xdot_{n-1})+G(\xdot_{n-1})\frac{1}{a}(x-x_{n-1})(\xdot_{n-1}-\xdot_{n-2}) \nn
+\frac{m}{2}(\frac{dx}{dt})^2 
-\frac{k}{2}(x-Vt)^2+\frac{K}{2a^2}(x-2x_{n-1}+x_{n-2})^2+I_n^0
,
\label{BK1}
\eea 
%%%%%%%%%%%%%%%%%%%%%%%%%%%
where $\xdot_n=d\xn(t)/dt$. $t$ is the time variable. 
$I_n^0$ is a constant term which does not depend on $x(t)$. 
We consider the system of $N$ particles (blocks) which distribute 
over the (1-dim) space $\{ y \}$. 
$m$ is the mass of one block. $k$ and $K$ are the spring-constants, 
the former is for the springs connecting the blocks, the latter is for 
the springs connecting to the moving (velocity $V$ ) 'plate'. 
The coordinate $y$ is taken to be periodic: $y\ra y+2L$. 
The particles (blocks) are moving around the equilibrium points 
$\{ P_n~|~n=1,2,\cdots,N-1,N \}$ where $P_N\equiv P_0$~(periodic boundary condition). 
The point $P_n$ is located at $y=y_n\equiv na$ ($Na=2L$) where $a$ is the 'lattice-spacing'.  
$N(=2L/a)$ is a huge number and the present system constitutes the statistical ensemble.  
The n-th particle's position at $t$, $\xn (t)$ (deviation 
from the equilibrium point $P_n$) is determined by the energy minimal 
principle $\del I_n(x)|_{x=\xn}=0$ with the pre-known movement of the (n-1)-th particle, $x_{n-1}(t)$, 
and that of the (n-2)-th, $x_{n-2}(t)$.
%*** BK2%%%%%%%%%%%%%%%%
\bea
-m\frac{d^2\xn}{dt^2} -F(\xdot_{n-1})+G(\xdot_{n-1})~\frac{\xdot_{n-1}-\xdot_{n-2}}{a}\nn
-k~(\xn-Vt)+\frac{K}{a^2}~(x_n-2x_{n-1}+x_{n-2})=0,
\label{BK2}
\eea 
%%%%%%%%%%%%%%%%%%%%%%%%%%%
where $0\leq t\leq \be$, and 
$F(\xdot_{n-1})$ and $G(\xdot_{n-1})$ are some functions of $\xdot_{n-1}$. 
This recursion relation determines the space-propagation. 
We assume the system is periodic in time:\ $t\ \ra\ t+\be$. 
This is Burridge-Knopoff model (Fig.\ref{Fig6BK}). 
(See the recent review article ref.\cite{KHK2012} for the use of BK model in the earthquake phenomena. )
G is newly introduced in the present paper. 

In the continuous space limit: 
%*** BK3%%%%%%%%%%%%%%%%
\bea
aN=2L\ (\mbox{fixed}),\ a\ \ra\ +0,\ N\ \ra\ \infty,\nn
y_n\ \ra\ y,\ x_n\ \ra\ x,\ (x_n-x_{n-1})/a\ \ra\ \pl x/\pl y,
\label{BK3}
\eea 
%%%%%%%%%%%%%%%%%%%%%%%%%%%
the step flow equation (\ref{BK2}) reduces to 
%*** BK4%%%%%%%%%%%%%%%%
\bea
-m\frac{\pl^2x}{\pl t^2}-F(\xdot)+G(\xdot)\frac{\pl^2x}{\pl y\pl t}
-k(x-Vt)+K\frac{\pl^2x}{\pl y^2}=0,                               \nn
x=x(t,y)\com\q \xdot=\frac{\pl x(t,y)}{\pl t}
\pr
\label{BK4}
\eea 
%%%%%%%%%%%%%%%%%%%%%%%%%%%
Note that, in the third term of the above equation, there appears the {\it velocity-gradient} 
$\frac{\pl^2x}{\pl y\pl t}$.  
(When the system slowly oscillates and G is expanded as $G(\xdot)=\eta+c_1\xdot+\cdots$, 
the first constant is the {\it viscosity}. )
The physical dimensions of the 
parameters in (\ref{BK4}) are listed as
%*** BK4w %%%%%%%%%%%%%%%%
\bea
[m]=\mbox{M},\ [k]=\mbox{M}\mbox{T}^{-2},\  
[V]=\mbox{L}\mbox{T}^{-1},\ [K]=\mbox{M}\mbox{L}^2\mbox{T}^{-2}
,
\label{BK4w}
\eea 
%%%%%%%%%%%%%%%%%%%%%%%%%%%
where we assume $[x]=[y]=$L, $[t]=$T and $[a]=$L. 

 Using the periodicity in time, we can read the {\it metric} ({\it geometry}) 
of this mechanical system. 
%*** BK5%%%%%%%%%%%%%%%%
\bea
{\Del s_n}^2\equiv 2a^2(I_n(\xn)-I_n^0)=\nn 
\{ -2\xn F(\xdot_{n-1}) +m{\xdot_n}^2 -k(\xn-Vt)^2  \}dy^2\nn 
-a\frac{\pl G(\xdot_{n-1})}{\pl t}~{\Del \xn}^2 + Ka^2~{\Del \vtil_n}^2\com\q dy\equiv a,\nn
\Del \xn\equiv \xn -x_{n-1},\ 
\frac{\xn-x_{n-1}}{a}\equiv \vtil_n,\ \vtil_n - \vtil_{n-1}=\Del \vtil_n
,
\label{BK5}
\eea 
%%%%%%%%%%%%%%%%%%%%%%%%%%%
where we assume $\Del \xdot_{n-1}=\Del \xdot_{n}$. 
$\vtil_n$ is the {\it longitudinal strain}. 
The full(complete) metric, $G_{IJ}$, is constructed from (\ref{BK5}) as
%*** BK5b%%%%%%%%%%%%%%%%
\bea
{\widetilde {ds}}^2= 
\{ -2x F(v) +mv^2 -k(x-Vt)^2  \} (dy^2-dt^2)\nn 
+ma^2{dv}^2 -a\frac{\pl G(v)}{\pl t}~{dx}^2 + Ka^2~(\frac{\pl v}{\pl y})^2{dt}^2\nn
=e_1 G_{IJ}(X)dX^I dX^J, e_1=K a^2~\mbox{or}~ ma^2V^2, v\equiv \xdot=\frac{\pl x}{\pl t}, \nn
(X^I)=(X^0,X^1,X^2,X^3)=(t/d_0,y/d_1,x/d_2,v/d_3),\nn
d_0=\sqrt{\frac{m}{k}}, d_1=V\sqrt{\frac{m}{k}}, d_2=\sqrt{\frac{K}{k}}, 
d_3=\sqrt{\frac{K}{m}}
,
\label{BK5b}
\eea 
%%%%%%%%%%%%%%%%%%%%%%%%%%%
where we use $d\vtil=d(\pl x/\pl y)=(\pl v/\pl y)dt$. 
$X^I$ are the dimensionless coordinates. 
Note that we have here done the natural replacement: 
$dy^2\ \ra\ -dt^2+dy^2$ and the addition of $ma^2{dv}^2$. 
(For the damped harmonic oscillator limit, $F=G=0, V=K=0, dy=0$, 
the line element reduces to $-m{dx}^2+kx^2{dt}^2+ma^2{dv}^2$\cite{SI1010ICSF}.  )
$G_{IJ}(X)$ is the metric in 4D manifold $(X^I)$. 
Note that both $d_1$ and $d_2$ have the same dimension of Length. 
$d_1$ (large scale) comes from $V$, whereas $d_2$ (small scale) from $K$. 
The 2 coordinates, $y$ and $x$, both describe the position in the space, 
but their roles are different.  
In the terminology of the general relativity, the former is the {\it general coordinate} 
and the latter the {\it local coordinate}. 
(The momentum coordinate $X^3=v/d_3$ is the "extra" coordinate in the holographic view.)\ 

\begin{figure}[h]
\includegraphics[bb= 0 0 248 116]{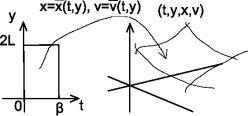}
\caption{
%****Figure7.eps*** 
{\it The two dimensional surface, (\ref{BK5c}),  in 4D space (t,y,x,v). }
        }
\label{Figure7}
\end{figure}

Now we take the following map from the 2D space $\{(t,y)|~0\leq t\leq \be,~0\leq y\leq 2L\}$ to 
the 4D space $(t, y, x, v)$. (The 4D space is called "target space" in the string-theory community. )
%*** BK5c%%%%%%%%%%%%%%%%
\bea
x=\xbar (t,y),\ v=\vbar(t,y),\nn
d\xbar=\frac{\pl\xbar}{\pl t}dt+\frac{\pl\xbar}{\pl y}dy,\ 
d\vbar=\frac{\pl\vbar}{\pl t}dt+\frac{\pl\vbar}{\pl y}dy
.
\label{BK5c}
\eea 
%%%%%%%%%%%%%%%%%%%%%%%%%%%
This map expresses a 2D {\it surface} in the 4D space (Fig.\ref{Figure7}). 
On the surface, the line element (\ref{BK5b}) 
reduces to
%*** BK5d%%%%%%%%%%%%%%%%
\bea
{\widetilde{ds}}^2\ -\mbox{on surface}\ra\  
e_1 g_{ij}(X)dX^i dX^j,                    %\ (X^0,~X^1)=(t/d_0,~y/d_1)
\ \ \ 
g_{00}=\nn
\frac{a^2}{e_1}
\left\{-H(\xbar,\vbar)+ma^2(\frac{\pl\vbar}{\pl t})^2
-\frac{\pl G}{\pl t}(\frac{\pl\xbar}{\pl t})^2+Ka^2(\frac{\pl\vbar}{\pl y})^2\right\},\nn
g_{01}=g_{10}=\frac{a^2\sqrt{m}}{{e_1}^{3/2}}
\left\{ ma^2 \frac{\pl\vbar}{\pl t}\frac{\pl\vbar}{\pl y}
         -\frac{\pl G}{\pl t}\frac{\pl\xbar}{\pl t}\frac{\pl\xbar}{\pl y} \right\},\nn
g_{11}=\frac{a^2}{e_1}
\left\{ H(\xbar,\vbar)+ma^2(\frac{\pl\vbar}{\pl y})^2 
-\frac{\pl G}{\pl t}(\frac{\pl\xbar}{\pl y})^2 \right\},\nn
H(\xbar,\vbar)\equiv -2\xbar F(\vbar)+m\vbar^2-k(\xbar-Vt)^2
,
\label{BK5d}
\eea 
%%%%%%%%%%%%%%%%%%%%%%%%%%%
where 
$\frac{\pl G}{\pl t}=\frac{dG(\vbar)}{d\vbar}\frac{\pl\vbar}{\pl t}$ and  
$i=0,1$. 
The number of blocks $N$ is a huge number, the present system statistically 
fluctuates. We regard the ensemble as the collection of possible surfaces $(\xbar(t,y),\vbar(t,y))$ in the 4D space $(t,y,x,v)$. The probability for each surface (system configuration) 
is specified by its area $A(\xbar,\vbar)$ as follows. 
Using the (dimensionless) surface area $A$, the partition function $\e^{-\be F}$ is given by
%*** se5x%%%%%%%%%%%%%%%%
\bea
A[\xbar(t,y),\vbar(t,y)]=\frac{1}{d_0 d_1}\int_0^\be dt\int_0^{2L}dy\sqrt{\det g_{ij}},\nn
\e^{-\be F}=\int 
\prod_{t,y}\Dcal \xbar(t,y)\Dcal \vbar(t,y)\e^{-\frac{1}{\al}A}
,
\label{se5x}
\eea 
%%%%%%%%%%%%%%%%%%%%%%%%%%%
where $\al$ is a dimensionless model parameter. 
$F$ is the free energy. 
$\al$ is determined by comparing the present result 
and the experimental data. 
%For example $\al=$ $\sqrt{K/m}(a^2/V^2)$, $\sqrt{K/k}\frac{a}{V}$ or $\sqrt{mK/k^2}$. 
The {\it minimum area surface}, which gives the main contribution to 
the above quantity, is given by the following equation.
%*** se5y%%%%%%%%%%%%%%%%
\bea
\frac{\pl A}{\pl\xbar(t,y)}=0\ ,\ \ \frac{\pl A}{\pl\vbar(t,y)}=0
.
\label{se5y}
\eea 
%%%%%%%%%%%%%%%%%%%%%%%%%%%

%%%%%%%%%%%%%%%%%%%%%%%%%%%%  Sec.4  %%%%%%%%%%%%%%%%%%%%%%%%%%%%%%%%%
%%%%                Conclusion                                    %%%%%%%%%%%%
%%%%%%%%%%%%%%%%%%%%%%%%%%%%%%%%%%%%%%%%%%%%%%%%%%%%%%%%%%%%%%%%%%%%%
%{\bf\large 4.\q Conclusion}\q
\section{ CONCLUSION  \label{conc}}
%***label***{conc}

We have treated two friction models: the spring-block model and Burridge-Knopoff model. 
Both are simple earthquake models. We have presented how to evaluate the 
statistical fluctuation effect. It is based on the geometry appearing in the system 
dynamics. In the text, the metric is obtained in (\ref{se1}) and (\ref{se2}) 
for the SB model and in (\ref{BK5b}) for the BK model.  

Multiple scales exist in both models. 
For the SB model, two length scales, one is from the natural length of the string $\lbar$ and 
the other from the external velocity $\Vbar$. For the BK model, three length scales exist. 
The one from the external velocity $V$, that from the spring constant $K$ and 
the block spacing $a$. 
As shown in the text, the use of dimensionless quantities clarifies the description. 
The multiple scales indicate the existence of the fruitful phases 
in the present statistical systems. 

As described in (\ref{SB2}) and (\ref{BK2}), the dissipative systems are treated by using the {\it minimal principle}. The difficulty of the {\it hysteresis} effect  (non-Markovian effect)  \cite{SI1010ICSF} is 
avoided in the present approach. These are the advantage of the discrete Morse flow method. %\cite{Kikuchi,Kikuchi2}. 
We do not use the ordinary time $t$, instead, exploit the step number $n$ ($t_n=nh$).  

We have presented several theoretical proposals for 
the statistical ensembles appearing in 
the friction phenomena. 
In order to select which one is the most appropriate, it is necessary to 
{\it numerically} evaluate the models with the proposed ensembles  
and compare the result with the real data 
appearing both in the natural phenomena and in the laboratory experiment.

\vspace*{5mm}
{\large Acknowledgment}
\vspace*{2mm}

The author thanks T. Hatano (Earthquake Research Inst., Univ. of Tokyo) for 
introducing ref.\cite{KHK2012} and the general discussion about earthquake. 
He also thanks H. Kawamura (Dep. of Earth and Space Science, Osaka Univ.) 
for the explanation about the numerical simulation of Burridge-Knopoff model.

%\end{multicols}



%% References


%%%%%%%%%%%%%%\bibliographystyle{elsarticle-num}
%%%%%%%%%%%%%%\bibliography{<your-bib-database>}






% \end{thebibliography}

\onecolumn
                 %%%%%%%%%%%%%%%%%%%%    <Fig.1  and 2     %%%%%%%%%%%%%%%%%%%
\vspace*{5mm}
                 %%%%%%%%%%%%%%%%%%%%    <Fig.3  and 4     %%%%%%%%%%%%%%%%%%%
                %%%%%%%%%%%%%%%%%%%%%%      Fig.3 and 4>   %%%%%%%%%%%%%%%%
%
%

                 %%%%%%%%%%%%%%%%%%%%    <Fig.5  and 6     %%%%%%%%%%%%%%%%%%%
                %%%%%%%%%%%%%%%%%%%%%%      Fig.5 and 6>   %%%%%%%%%%%%%%%%
%
%

                 %%%%%%%%%%%%%%%%%%%%    <Fig.7     %%%%%%%%%%%%%%%%%%%
                %%%%%%%%%%%%%%%%%%%%%%      Fig.7>   %%%%%%%%%%%%%%%%

\end{document}